\documentclass[preprint]{aastex631}

\hypersetup{linkcolor=red,citecolor=cyan,filecolor=blue,urlcolor=magenta}
\usepackage{subfigure}

\usepackage{color}
\shorttitle{NVSS J171822+423948: a possible neutrino emitter}
\shortauthors{Jiang et al.}

\begin{document}

\title{Awakening of A Blazar at Redshift 2.7 Temporally Coincident with Arrival of Cospatial Neutrino Event IceCube-201221A}

\correspondingauthor{Neng-Hui Liao; Ting-Gui Wang}
\email{nhliao@gzu.edu.cn;twang@ustc.edu.cn}

\author{Xiong Jiang}
\affiliation{Department of Physics and Astronomy, College of Physics, Guizhou University, Guiyang 550025, China}

\author[0000-0001-6614-3344]{Neng-Hui Liao}
\affiliation{Department of Physics and Astronomy, College of Physics, Guizhou University, Guiyang 550025, China}

\author[0000-0003-4225-5442]{Yi-Bo Wang}
\affiliation{School of Astronomy and Space Science, University of Science and Technology of China, Hefei, Anhui 230026, People’s Republic of China}

\author[0000-0003-1721-151X]{Rui Xue}
\affiliation{Department of Physics, Zhejiang Normal University, Jinhua 321004, China}

\author[0000-0002-7152-3621]{Ning Jiang}
\affiliation{School of Astronomy and Space Science, University of Science and Technology of China, Hefei, Anhui 230026, People’s Republic of China}
\affiliation{Key Laboratory for Research in Galaxies and Cosmology, Department of Astronomy, The University of Science and Technology of China, Chinese Academy of Sciences, Hefei, Anhui 230026, People’s Republic of China}

\author[0000-0002-1517-6792]{Ting-Gui Wang}
\affiliation{School of Astronomy and Space Science, University of Science and Technology of China, Hefei, Anhui 230026, People’s Republic of China}
\affiliation{Key Laboratory for Research in Galaxies and Cosmology, Department of Astronomy, The University of Science and Technology of China, Chinese Academy of Sciences, Hefei, Anhui 230026, People’s Republic of China}

\begin{abstract}
We report on multiwavelength studies of a blazar NVSS J171822+423948, which is identified as the low-energy counterpart of 4FGL J1718.5+4237, the unique $\gamma$-ray source known to be cospatial with the IceCube neutrino event IC-201221A. After a 12-year long quiescent period undetected by Fermi-LAT, $\gamma$-ray activities with a tenfold flux increase emerge soon (a few tens of days) after arrival of the neutrino. Associated optical flares in the ZTF $g$, $r$, and $i$ bands are observed together with elevated WISE infrared fluxes. Synchronized variations suggest that both the $\gamma$-ray emission and the neutrino event are connected to the blazar. Furthermore, the optical spectrum reveals emission lines at a redshift $z$ = 2.68 $\pm$ 0.01. Thus, it is the first candidate for a neutrino-emitting blazar at the redshift above 2. Discussions of theoretical constraints of neutrino production and comparisons with other candidates are presented.  
\end{abstract}

\keywords{neutrino astronomy; active galactic nuclei; Gamma-ray sources}

\section{Introduction} \label{sec:intro}

The IceCube Neutrino Observatory has detected high-energy astrophysical neutrinos \citep[e.g.,][]{2013Sci...342E...1I,2015PhRvL.115h1102A,2023Sci...380.1338I}, ushering in a new era in neutrino astronomy. The neutrino signal is compatible with an isotropic distribution in the sky, suggesting that a substantial fraction of the observed neutrinos are of extragalactic origin. Neutrinos interact weakly with matter and travel undeflected in magnetic fields, providing key insights into extreme cosmic environments that are otherwise opaque \citep{2018AdSpR..62.2902A}. Interactions involving photohadronic ($p\gamma$) or hadronuclear ($pp$) processes are responsible for the generation of neutrinos, while accompanying $\gamma$-ray radiations are also anticipated \citep{2022arXiv220200694H}. Besides routine leptonic radiations, hadronic cascades also produce electromagnetic emissions at lower energies. Hence, potential connections between electromagnetic emissions and neutrino events are suggested. Various types of extragalactic $\gamma$-ray emitters are proposed as potential neutrino sources, such as active galactic nucleus (AGN) jets \citep[e.g.,][]{2013ApJ...768...54B,2014PhRvD..89l3005B}, $\gamma$-ray bursts \citep[e.g.,][]{1995PhRvL..75..386W}, tidal disruption events \citep{2021NatAs...5..510S,2022PhRvL.128v1101R,2023ApJ...953L..12J}, starburst galaxies \citep[e.g.,][]{2006JCAP...05..003L} as well as galaxy clusters \citep[e.g.,][]{2008ApJ...689L.105M}. Intriguingly, evidence for neutrino emission from a nearby Seyfert galaxy NGC 1068 has been revealed \citep{2022Sci...378..538I}. However, because of the relatively large angular uncertainties of neutrinos, the origin of extragalactic astrophysical neutrinos remains unresolved.

In addition to spatial information, temporal information is also crucial for determining the origin of neutrinos. IceCube has detected a 0.3 PeV muon neutrino (IC-170922A) cospatial with a blazar TXS 0506+056 ($z$ = 0.3365, \citealt{2018ApJ...854L..32P}), at the time when the blazar underwent unprecedented multi-wavelength flares, yielding a 3$\sigma$ significance for the correlation of observed multimessenger radiations \citep{2018Sci...361.1378I}. Blazars are a rare subclass of AGNs that includes flat-spectrum radio quasars (FSRQs) and BL Lacertae objects (BL Lacs), of which strong relativistic jets are well aligned with our light of sight \citep{2019ARA&A..57..467B}. Blazars, which are the dominant population in the extragalactic $\gamma$-ray sky, exhibit overwhelming and highly variable Doppler-boosted nonthermal jet emissions that cover a wide range of energy regimes from radio to very high-energy $\gamma$-rays \citep{2016ARA&A..54..725M}. There are a growing number of cases in which both spatial and temporal coincidences have been observed between flaring blazars and incoming neutrinos \citep[e.g.,][]{2016NatPh..12..807K, 2019ApJ...880..103G, 2020ApJ...893..162F, 2020A&A...640L...4G, 2022ApJ...932L..25L, 2023MNRAS.519.1396S}. However, such circumstances are still rather rare. Furthermore, from an observational perspective, the connections between electromagnetic emissions and neutrinos are rather intricate. Electromagnetic activities were not observed during neutrino flaring of TXS 0506+056 in 2014 - 2015 \citep{2018Sci...361..147I}. However, one of the brightest $\gamma$-ray blazars, PKS 1502+106, lies within the localization uncertainty region of the neutrino event IC-190730A \citep[e.g.,][]{2021ApJ...912...54R}. It was in a quiescent state from optical emission to $\gamma$ rays, but with high flux-level radio emission, when the neutrino arrived \citep{2021ApJ...911L..18K}.

On 2020 December 21, a 0.174~PeV track-like neutrino event (IceCube-201221A, hereafter IC-201221A) with a 56.4\% probability of being an astrophysical origin (i.e., {\tt Gold} alert streams) with the arrival direction given as R.A. $261.69^{+2.29}_{-2.50}$$^{\circ}$ and decl. $41.81^{+1.29}_{-1.20}$$^{\circ}$ was reported \citep{2020GCN.29102....1I}. Subsequent electromagnetic observations were triggered, but no corresponding activities were reported \citep[e.g.,][]{2020GCN.29135....1G, 2020GCN.29151....1F}. However, the fourth data release of the Fermi$\gamma$-ray source catalog (4FGL-DR4, \citealt{2023arXiv230712546B}) includes a listing for a $\gamma$-ray source, 4FGL J1718.5+4237, in this direction. However, no corresponding low-energy counterpart is provided. Given the absence of a known source in the 4FGL-DR3 catalog \citep{2022ApJS..260...53A}, the detection of 4FGL J1718.5+4237 is likely attributed to its recent flaring activity, which suggests a possible temporal correlation with IC-201221A. In light of these observational findings, we thoroughly examine 4FGL J1718.5+4237 in this letter, including the search for its low-energy counterpart, and explore its potential association with the neutrino event IC-201221A. Analyses of multiwavelength data, including an optical spectroscopic observation by Hale 5 m telescope, along with results, are presented in Section 2; while discussions and a short summary are given in Section 3. We adopt a $\Lambda$CDM cosmology with $ \Omega_{M} $ = 0.32, $ \Omega_{\Lambda} $ = 0.68, and a Hubble constant of $H_{0}$ = 67 km$^{-1}$ s$^{-1}$ Mpc$^{-1}$ \citep{2014A&A...571A..16P}.

\section{Data Analysis and Results} \label{sec:data}
\subsection{Fermi-LAT Data} \label{subsec:tables}
We analyzed the first 14.4-yr (i.e., from 2008 August 04 to 2023 January 01) Fermi-LAT Pass 8 data ({\tt evclass} = 128 and {\tt evtype} = 3) between energy 100 MeV and 500 GeV. We used {\tt Fermitools} software version 2.0.8 and the accompanying {\tt Fermitools-data} version 0.18. A region of interest (ROI), centered on 4FGL J1718.5+4237, with a 10-degree radius was considered. In order to avoid significant contamination from earth limb, a zenith angle cut ($<$ 90$ ^{\circ}$) was adopted, so as the quality-filter cuts (i.e., {\tt DATA\_QUAL==1 \&\& LAT\_CONFIG==1}). {\tt Unbinned} likelihood analysis was carried out using {\tt gtlike} task to extract the flux and spectrum of $ \gamma$ rays. The significance of detecting a $\gamma$-ray source was determined by the test statistic (TS = -2ln($ L_{0}$/$L$), \citealt{1996ApJ...461..396M}). $L$ and $L_{0}$ denote the maximum likelihood value for the model with and without the target source, respectively. The parameters of the background sources within the ROI as well as normalizations of the two diffuse templates (i. e., {\tt gll\_iem\_v07} and {\tt iso\_P8R3\_SOURCE\_V3\_v01}) were left free, while other parameters were fixed to values listed in 4FGL-DR4. We examined whether new $\gamma$-ray sources emerge (i.e., TS $ >$ 25) by analyzing subsequently generated residual TS maps. If so, we then reran the likelihood fitting process with the updated background model. In the process of temporal analysis, low-intensity background sources (TS $<$ 10) were excluded from the model file. Additionally, the {\tt pyLikelihood UpperLimits} tool was employed to compute a 95\% confidence level (C. L.) upper limit instead of a flux, in cases of tentative detections.

First, a complete analysis of the entire dataset was performed, from which the detection of a significant $\gamma$-ray source (TS = 60) in this direction was confirmed. The average flux of $ \gamma$ rays is given as (4.05 ± 1.07) $\times$ 10$ ^{-9}$ ph cm$^{-2}$ s$^{-1}$, along with a relatively soft spectrum ($\Gamma$ = 2.57 $\pm$ 0.19, where $\Gamma$ is the index of the power law function, $ dN/dE$ $\propto$ ${E}^{-\Gamma}$). The optimized location was constrained as R.A. $259.642^{\circ}$, decl. $42.6215^{\circ}$, with a 95\% C.~L. error radius of $ 5\arcmin$. These results are consistent with those listed in 4FGL-DR4. Within the radius of localization error, two radio sources, NVSS J171822+423948 (hereafter cand. \uppercase\expandafter{\romannumeral1}) and NVSS J171853+423658 (hereafter cand. \uppercase\expandafter{\romannumeral2}), are found. Therefore, the association relationship cannot be determined by spatial information alone. Investigations of two candidates were carried out. The former has been identified as a flat spectral radio source \citep{2003MNRAS.341....1M}, while for the latter, data only from the NRAO VLA sky survey (NVSS) are accessible \citep{1998AJ....115.1693C}. Sloan Digital Sky Survey (SDSS) reveals a possible dropout in the $u$ band for cand. \uppercase\expandafter{\romannumeral1} \citep{2020ApJS..249....3A}, suggesting that it may be a high redshift quasar. Based on the SDSS $g$ band magnitude and radio flux density at 5~GHz, radio-loudness of cand. \uppercase\expandafter{\romannumeral1} is given as high as $\sim$ 1500 \citep{1989AJ.....98.1195K}, indicative of a blazar candidate.

\begin{figure*}
    \centering
    \includegraphics[scale=0.4]{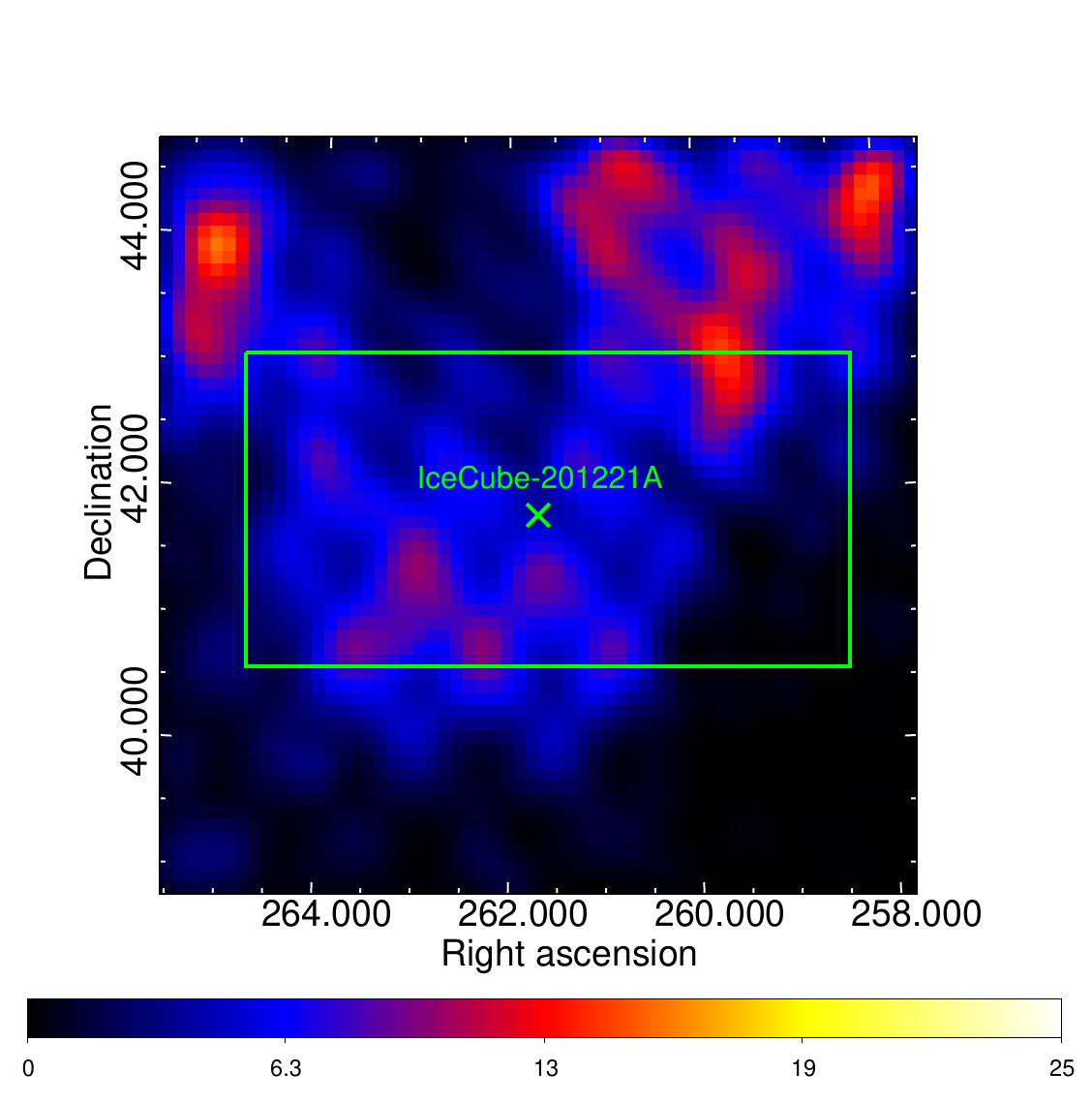}
    \includegraphics[scale=0.4]{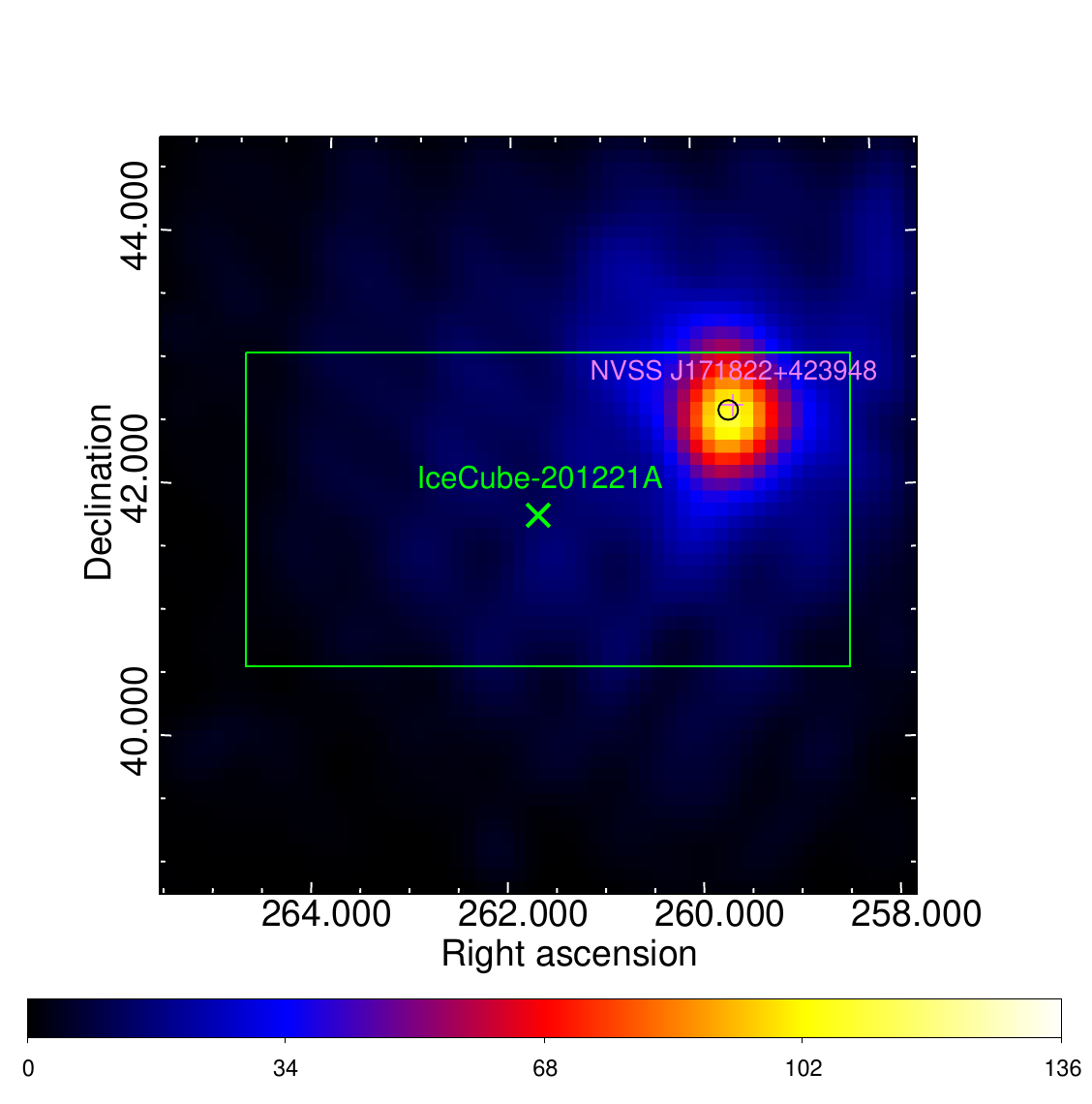}
    \includegraphics[scale=0.6]{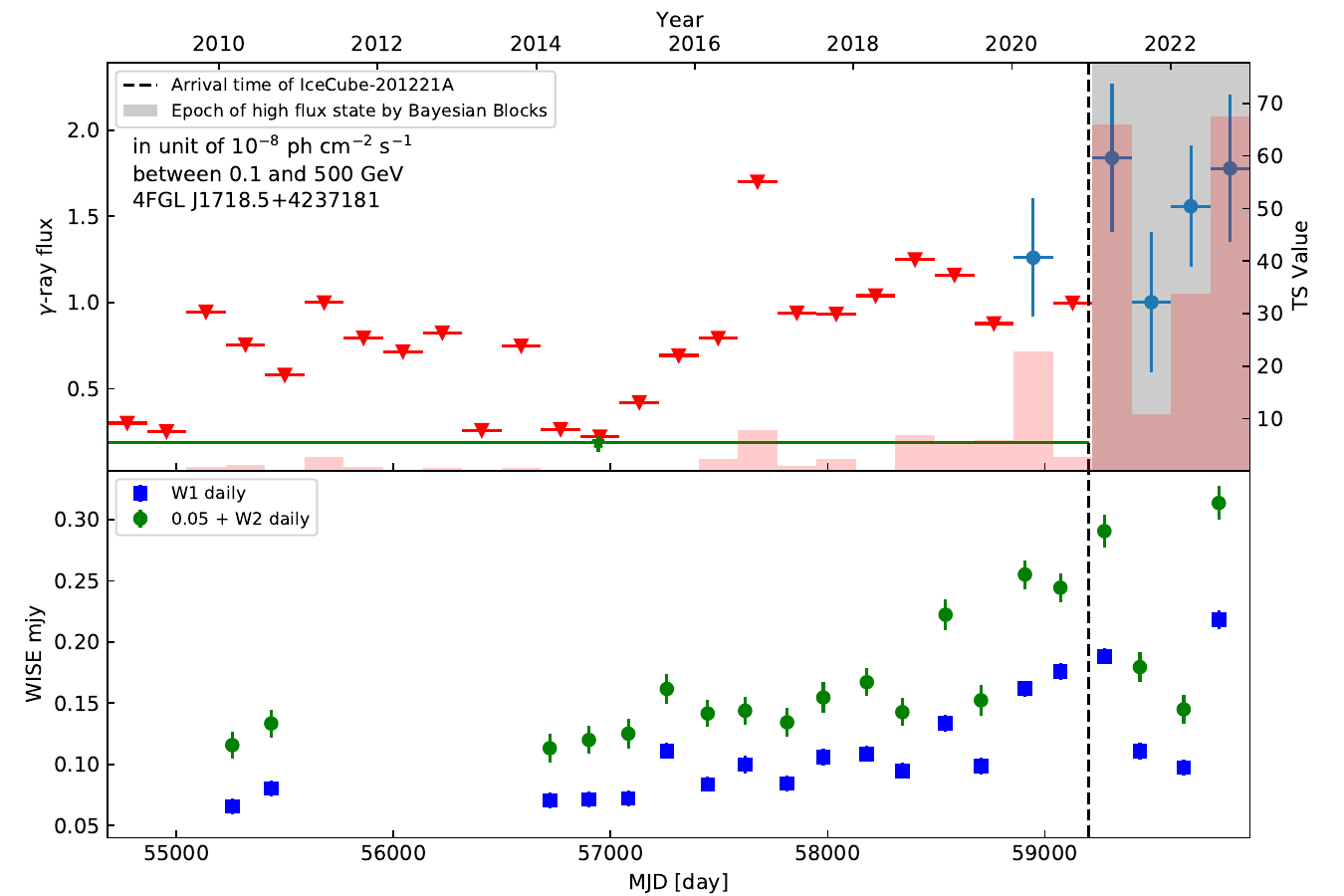}
    \caption{{\bf Upper panels:} smoothed $ \gamma$-ray TS maps (6$ \degr\times 6\degr$ scale with 0.1$ \degr$ per pixel, 4FGL J1718.5+4237 not include in the background model). The left one is based on the Fermi-LAT data between MJD 54683 and MJD 59204 (prior to arrival of the neutrino). While the right one corresponds to the Fermi-LAT data for rest of the time. The green X-shaped symbol and rectangle represent optimized position and positional uncertainty of the neutrino, respectively. The blue circle is the 95\% C.L. $\gamma$-ray localization error radius of 4FGL J1718.5+4237, and the violet cross marks radio location of cand. I. {\bf Bottom panel}: $\gamma$-ray light curves of 4FGL J1718.5+4237 and infrared ones of cand. I. In the former, blue circles and red triangles correspond to flux estimations and upper limits, along with TS values displayed as red bars. The gray shaded area represents an epoch of high $ \gamma$-ray flux state obtained by Bayesian blocks. The black vertical dashed line across the light curves marks the arrival time of IC-201221A.
}
    \label{glc}
\end{figure*}

A half-year time-bin $ \gamma$ -ray light curve was extracted to investigate its temporal behavior, shown in Figure \ref{glc}. A distinct feature of the light curve is the significant flux variation on a long timescale. Adopting ``variability index'' test \citep{2012ApJS..199...31N}, the significance of the variability is given as $> 5\sigma$. Considering the relatively limited angular resolution of Fermi-LAT for sub-GeV photons, light curves from nearby background sources were also extracted. Since no similar variations were found to those of the target, therefore, the variation of the target is suggested to be intrinsic rather than artificially caused by the neighbors. Furthermore, the Bayesian Block analysis \citep{2013ApJ...764..167S} marks a two-year epoch (i.e., last four time bins) in a high-flux state between MJD 59219 and 59945. A specific analysis yields a significant $\gamma$ ray source (TS = 136). The corresponding photon flux is (1.4 $\pm$ 0.24) $\times$ 10$ ^{-8}$ ph cm$^{-2}$ s$^{-1}$, with a slightly harder spectrum ($\Gamma$ = 2.43 $\pm$ 0.11) than the value derived from the analysis of the entire dataset. No significant spectral curvature is found by sophisticated spectral templates. By utilizing the Fermi {\tt gtsrcprob} tool, the most energetic photon is at approximately 10 GeV, while the majority of high-energy $\gamma$-ray photons possess an energy of around 2 GeV. The localization analysis provides an optimized position of  R.A. $= 259.64^{\circ}$ and decl. $= 42.6165^{\circ}$, with a 95\% C.~L. error radius of approximately $4\arcmin$. Both candidates remain within the error radius. This source had been maintained in a low-flux state for the 12 years since the beginning of Fermi-LAT observation. Combining the low-state data produces only a tentative detection (TS = 15) of a flux (1.89 $\pm$ 0.58)$\times$ 10$ ^{-9}$ ph cm$^{-2}$ s$^{-1}$, which is approximately seven times lower than the peak value.

We extracted a zoomed-in 2-month time-bin light curve, focusing on the epoch of the high-flux state (see Figure \ref{mlc}). The $\gamma$-ray activity consists of two sub-flares, separated by a one-year long quiescent state. The first one lasts about four months (TS = 54), between MJD 59214 and 59336. The duration of the other flare is also several months (TS = 113). Individual analyses of data for these two subflares reveal a similar flux level, $\simeq$ 2 $\times$ 10$ ^{-8}$ ph cm$^{-2}$ s$^{-1}$, approximately one order of magnitude of the flux level extracted from the first 12-year Fermi-LAT data. The two subflares also share a similar spectrum (i.e., $\Gamma \simeq$ 2.3). In addition, localization analyses provide similar results so that the two radio sources fall into the $\gamma$-ray radii. Weekly time-bin $ \gamma$ -ray light curves were further extracted, but no sign of variability at a timescale of days is found.

\subsection{Optical Data}
\subsubsection{Hale P200 spectroscopic observation}
The spectroscopic observation of Cand. \uppercase\expandafter{\romannumeral1} was performed at MJD 60223, using the double spectrograph, with dichroic D55, mounted on the Hale 200-inch telescope at Palomar Observatory (P200, \citealt{1982PASP...94..586O}). The total exposure time for the observation is 2400 s.  To match the seeing conditions, we chosed a slit width of 1.5$\rm \arcsec$. Data were reduced with {\tt Pypeit} (\citealt{2020JOSS....5.2308P}). We conducted flux calibration using a standard star on the same night. Galactic extinctions were corrected \citep{2011ApJ...737..103S}. A simple power-law model was used to fit the continuum component after excluding the wavelength windows of the emission lines, the telluric region, and the blueward region of the peak of $\rm Ly\alpha$.

\begin{figure*}[htb]
\centering
\begin{minipage}{0.7\textwidth}
\centering{\includegraphics[angle=0,width=1.0\textwidth]{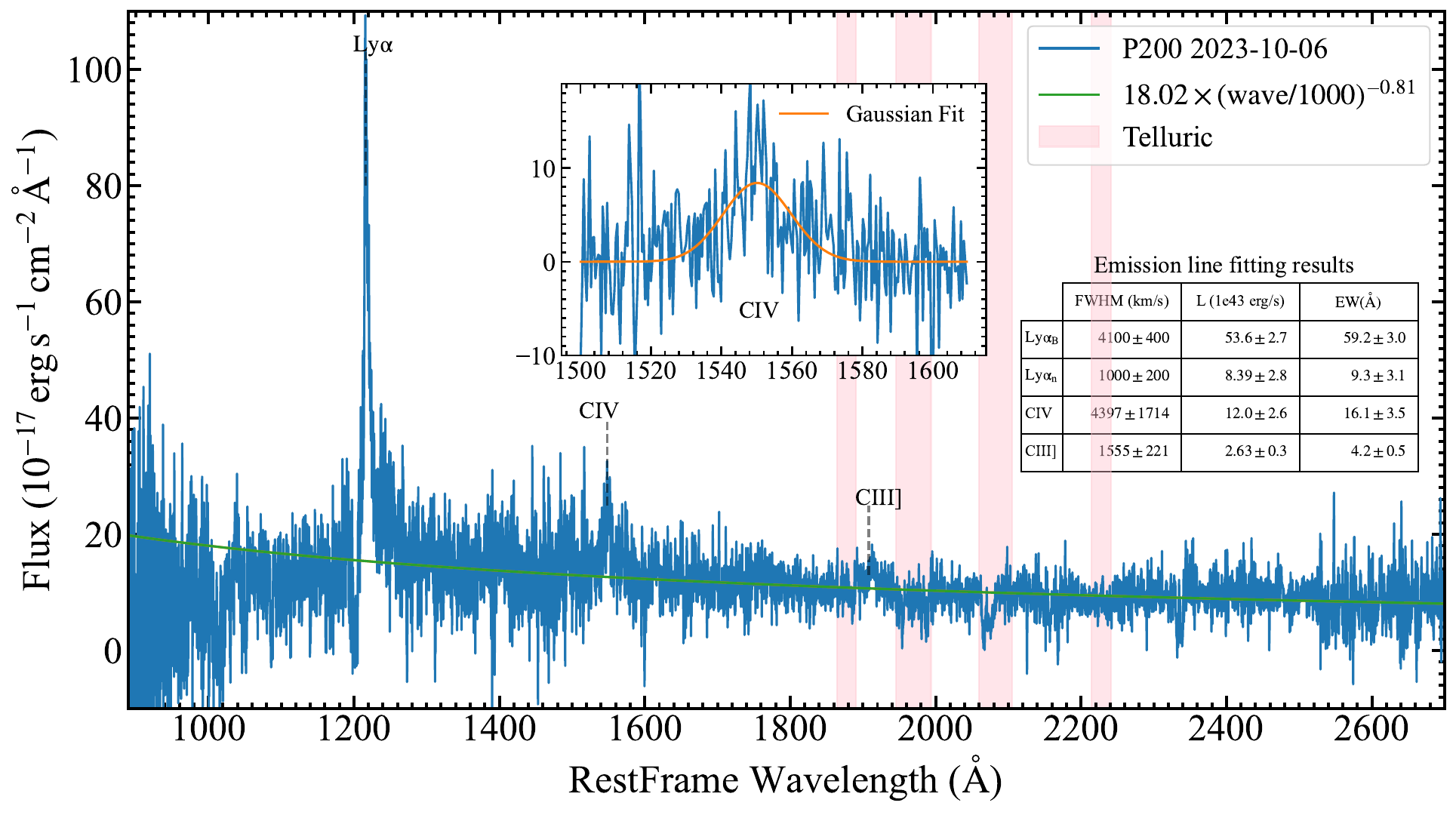}}
\end{minipage}
\caption{P200 Spectrum. The spectrum is depicted in blue, the power-law continuum in green, and the Gaussian fitting for the CIV emission lines in orange. The red shadows represent the positions of the dichroic or the uncorrected telluric. Fitting results for the emission lines, including  broad and narrow components of $\rm Ly\alpha$ line, as well as $\rm C\,\textsc{iv}$ and $\rm C\,\textsc{iii]}$ lines, are listed.}
\label{spec}
\end{figure*}

The spectrum shows strong $\rm Ly\alpha$, moderately weak $\rm C\,\textsc{iv}$, and faint $\rm C\,\textsc{iii}$ emission lines, which give a redshift of 2.68$\pm$ 0.01, as shown in Figure \ref{spec}. $\rm C\,\textsc{iv}$ emission line was fitted with a single Gaussian. $\rm FWHM_{CIV}$ is given as $\sim 4397\pm1714$ km $\rm s^{-1}$, of which the uncertainty is derived by Monte-Carlo simulations, also listed in Figure \ref{spec}. Based on the estimated $\rm \lambda\,L_{1350} \sim 1.12\times10^{46}\,erg\,s^{-1}$ and the bolometeric correction factor $\rm BC_{1350} =3.81$ \citep{2011ApJS..194...45S}, the bolometeric luminosity reaches to $\rm L_{bol, disk}\sim4.28\times10^{46}$ erg $\rm s^{-1}$. Then the mass of the SMBH can be inferred as $\rm log(M_{BH}/M_{\sun}) \sim 9$ \citep{2006ApJ...641..689V}. Therefore, the Eddington ratio $\rm r_{Edd}\sim$ 0.3 is suggested. Although peaking frequency of the synchrotron emission for FSRQs is typically at sub-mm/infrared domain, temporal jet contribution in optical bands may not be negligible, when fast and large-amplitude optical variations associated with jet activities are exhibited. In order to investigate influence of the jet emission in our optical spectrum, based on the relationship between $\rm L_{C\,\textsc{iv}}$ and $\rm \lambda\,L_{1350} $ of SDSS quasars \citep{2011ApJS..194...45S}, $\rm \lambda\,L_{1350}$ is deduced from the $\rm C\,\textsc{iv}$ line luminosity, which is consistent with that directly obtained from the continuum spectral component. Therefore, jet is likely at quiescent state at the time of the spectroscopic observation and its contribution in optical bands is not significant then. In addition, a description of archival SDSS measurements (u-band data excluded) that correspond to a low optical flux level (see Figure \ref{sed}), by a multi-temperature radial profile \citep{1973A&A....24..337S}, has been performed. A similar value of $\rm L_{bol, disk}$ $\sim 3\times10^{46}$ erg $\rm s^{-1}$ is yielded, compared with the result from the optical spectrum.

\subsubsection{Zwicky Transient Facility Light-curve Data}
We collected the latest Zwicky Transient Facility (ZTF) data \citep{2019PASP..131a8003M,ptf}\footnote{\url{https://www.ztf.caltech.edu/ztf-public-releases.html}}. Based on coadded reference images in the $g$, $r$, and $i$ bands, magnitudes within 5\arcsec~of Cand. \uppercase\expandafter{\romannumeral1} were extracted to construct optical light curves, shown in Figure \ref{mlc}. Only frames satisfied with {\tt catflag = 0} and {\tt chi < 4} were selected. Optical activities in multi-epoch, for instance, around MJD 58600, 59300 and 59700, are exhibited. The variability amplitudes in all three bands are approximately one magnitude. For Cand. \uppercase\expandafter{\romannumeral2}, no significant excesses against the background emission were found from the reference images in all three bands. 

\subsection{WISE data}
We investigated the infrared counterparts of the two candidates. An infrared source, WISE J171822.76+423945.1, is clearly detected at the position of Cand. \uppercase\expandafter{\romannumeral1}, while no sources were found within 10\arcsec\ of the radio position of Candidate \uppercase\expandafter{\romannumeral2}, according to the AllWISE Source Catalog \citep{2010AJ....140.1868W,allwise}. The \emph{W1} ($3.4~\mu$m) and \emph{W2} ($4.6~\mu$m) magnitudes given by the catalog are 16.587 $\pm$ 0.059 and 16.152 $\pm$ 0.123, respectively. We first attempted to construct the long-term light curves after excluding bad exposures and binning the data within each epoch (within one day) following~\citet{2021ApJS..252...32J}. Although with larger errors (typically $\simeq$ 0.2 mag), there are two obvious flux maxima at MJD 59274 (i.e. \emph{W1} = 15.412 $\pm$ 0.294 mag and \emph{W2} = 14.439 $\pm$ 0.317 mag) and MJD 59804 (i.e. \emph{W1} = 15.297 $\pm$ 0.166 mag and \emph{W2} = 14.411 $\pm$ 0.26 mag). To confirm this and obtain more accurate measurements, we also attempted to perform PSF photometry on the differential images of time-resolved WISE/NEOWISE coadds~\citep{2018AJ....156...69M} with the first epoch as a reference following~\citet{2021ApJ...911...31J}. Finally, we adopted the fluxes measured by subtraction of the image with the reference flux added (see Figure~\ref{glc}).

\subsection{Implications of multi-messenger observations}

\begin{figure*}
    \centering
    \includegraphics[scale=0.8]{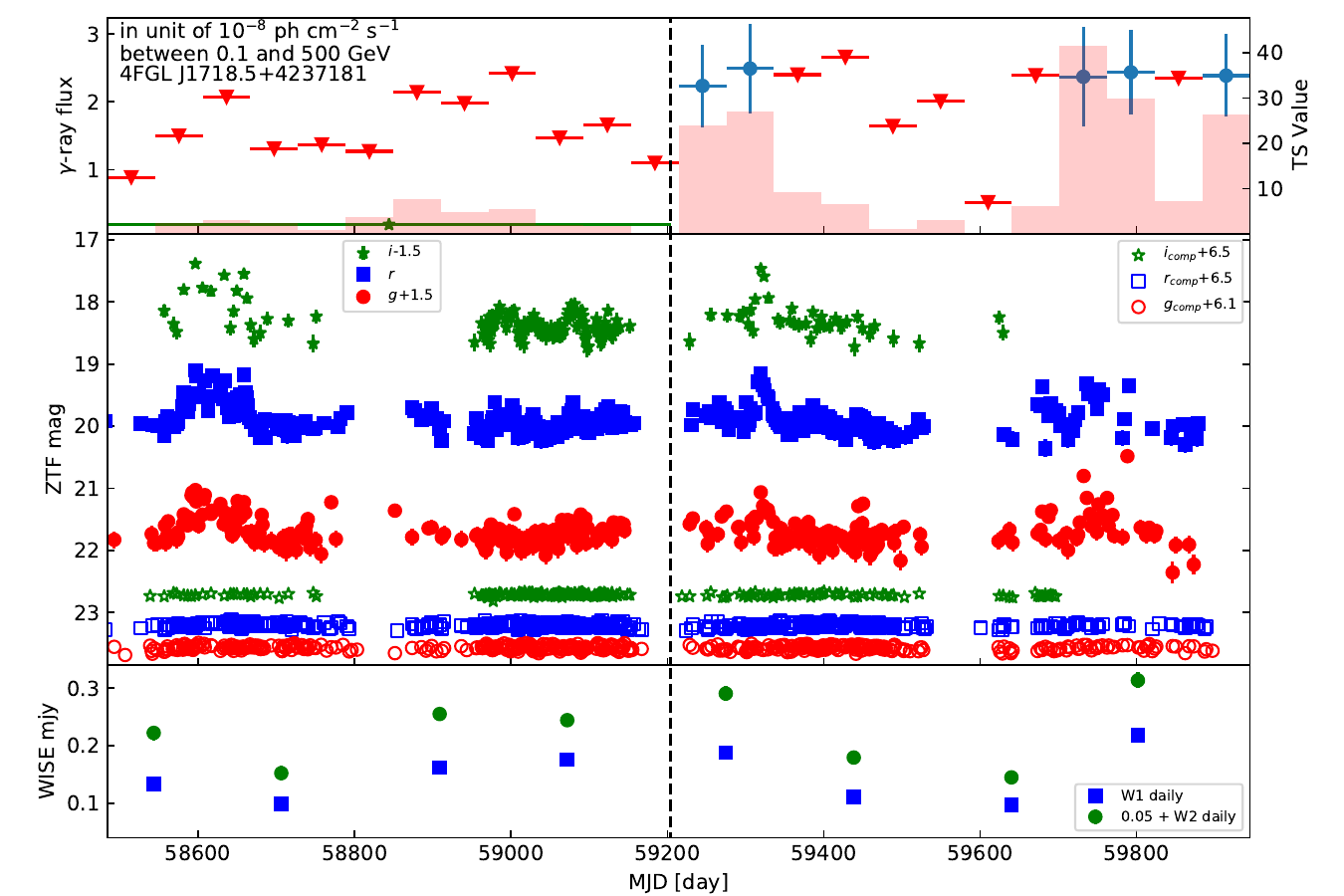}
    \caption{Zoomed-in multiwavelength light curves. {\bf Upper panel:} Two-month time bin light curve $\gamma$-ray light curves of 4FGL J1718.5+4237. {\bf Middle panel:} ZTF light curves of cand. I, solid markers represent the magnitudes of the target, while hollow ones correspond to the average values of the comparison stars in the same field. {\bf Bottom panel:} WISE light curves of cand. I. A black vertical dashed line across all panels marks the arrival time of the neutrino.}
    \label{mlc}
\end{figure*}

Blazars are well known for significant broadband variability, on timescales from minutes to years \citep{1995ARA&A..33..163W,1997ARA&A..35..445U}. A remarkable observational feature of FSRQs is that correlated optical/infrared and $\gamma$-ray flares are frequently detected \citep[e.g.,][]{2010Natur.463..919A}. Since the angular resolution of Fermi-LAT (typically $\simeq$ 0.1$^{\circ}$) is limited compared to those of the optical and radio bands ($\lesssim$ 3\arcsec), the observed correlated emissions are essential to determine the association relationship between $\gamma$-ray source and its low-energy counterpart. For the two radio candidates falling into the localization uncertainty area of 4FGL J1718.5+4237, optical and infrared temporal behaviors have been investigated here. Significant flux variations of cand. I have been detected, while no such evidence is found for the other one. Furthermore, contemporaneous brightenings between optical/infrared emissions of cand. I and $\gamma$-ray emissions of 4FGL J1718.5+4237 are revealed. As shown in Figure \ref{mlc}, corresponding to a $\gamma$-ray activity around MJD~59300, distinct optical flares in all three independent bands are presented. Meanwhile, its infrared emission is at high flux state. Similar behaviors are found for another $\gamma$-ray activity MJD~59800, though there is no available $i$-band observations then. Based on these observed features, in conclusion, cand. I (i.e., NVSS J171822+423948) is suggested as the low-energy counterpart of 4FGL J1718.5+4237. In the latest incremental version of the fourth catalog of AGNs detected by Fermi-LAT \citep{2022ApJS..263...24A}, only 29 known $\gamma$-ray emitting blazars beyond the redshift of 2.6 are embraced, among more than three thousand sources. Our optical spectroscopic observation reveals the redshift of cand. I as $z$ = 2.68 $\pm$ 0.01, and hence it is noticed as one of the rare high redshift $\gamma$-ray emitting blazars, which are crucial for studying relativistic jets and mass growth of the central SMBHs at the early epoch of the evolution of the universe \citep[e.g.,][]{2010A&ARv..18..279V,2017ApJ...837L...5A,2018ApJ...865L..17L}.

The contemporaneous brightening of cand. I suggests that its optical and $\gamma$-ray emission is likely from the same radiation region. Although no sign of fast variability in $\gamma$-ray is found, optical flares peaking at MJD~59319 put a constraint on the radiation region. For instance, the $i$-band flux increases 1.1 mag within 10 days, giving a doubling time, in cosmic rest frame, $\tau_{doub}=\Delta t\times {\rm ln}2/{\rm ln}(F_{1}/F_{2})/(1+z) \simeq 1.9$~day. A similar value is obtained from the flux variations in the $r$ band, but the timescale in $g$-band is relatively longer. Hence, adopting a routine Doppler factor ($\delta$) value of 10 \citep{2017MNRAS.466.4625L}, radius of the radiation region is limited to $r \lesssim \tau_{doub}c\delta \simeq$ 0.02 pc. Assuming a conical jet geometry, the location of the radiation region is constrained to $R_{loc} \sim \delta r \sim$ 0.2 pc, while the radius of the broad line region (BLR) is also suggested as $\sim$ 0.2 pc based on the detected $\rm \lambda\,L_{1350}$ \citep{2019ApJ...887...38G}. Therefore, the radiation region is likely embedded within the radiation of BLR. Note that the observational cadence of ZTF is several days, so potential intraday optical variability could be missed. In fact, fast $\gamma$-ray variability for blazars of $z \gtrsim$ 3 with a timescale down to a few hours has been reported \citep{2018ApJ...853..159L}. Meanwhile, intraday infrared variability has also been found in one of these sources \citep{2019ApJ...879L...9L}. All of these pieces of evidence indicate the presence of a compact dissipation region.

Benefiting from multi-messenger observations, the connection of the flaring blazar and the neutrino event is investigated. Firstly, cand. I is the only known $\gamma$-ray source spatially coincident with IC-201221A. However, due to its relatively large localization uncertainty, relying solely on spatial coincidence is not sufficient to claim a physical connection. Monte-Carlo simulations have been performed to calculate a probability of chance. The central location of IC-201221A is randomized in the sky beyond the Galactic plane with a varying right ascension value. Since the sensitivity of IceCube is strongly dependent on sky declination, the declination value of the neutrino is frozen during the simulation. Meanwhile, the size of the angular uncertainty box is also fixed. After $\rm 10^{4}$ simulations, the chance probability is derived as $p = \frac{M + 1}{N + 1}$ $\simeq$ 1, where N is the number of simulations, M is the number of 4FGL-DR4 sources \citep{2023arXiv230712546B} that fall into the simulated uncertainty box. Obviously, additional temporal information is needed. As shown in Figure \ref{glc}, incoming of the neutrino seems to switch on $\gamma$-ray emission of the blazar. It maintains to be in a quiescent state lasting for more than 12 years beneath the detection limit of Fermi-LAT. However, it becomes detectable soon after the arrival of IC-201221A, roughly several tens of days. The amplitude of variability between different flux levels is one order of magnitude. Since the yearly time bin $\gamma$-ray light curve for each 4FGL-DR4 source is available\footnote{\url{https://fermi.gsfc.nasa.gov/ssc/data/access/lat/14yr\_catalog/4FGL-DR4\_LcPlots\_v32.tgz}}, the simulations are re-performed including temporal information. At this time, only sources exhibiting a clear flux increase in the 13th time bin (i.e., three times of the 14-yr averaged flux) are considered. The probability of chance reduces to $\sim$ 0.03 from $\simeq$ 1.  A similar case in which the emergence of a cospatial transient $\gamma$-ray source coincides temporally with neutrinos detection is reported for GB6 J2113+1121/IC-191001A \citep{2022ApJ...932L..25L}. Apart from a long period in quiet state (i.e., over 8-yr), $\gamma$-ray flux enhancement about five times of TXS 0506 + 056 in long-term are found, corresponding to incoming of IC-170922A \citep{2018Sci...361.1378I}. Furthermore, in a multiwavelength perspective, strong optical flares with peaking time 115 days later of arrival time of the neutrino (i.e., MJD 59204) are detected, although prior optical activities are exhibited, see Figure \ref{mlc}. In addition, accompanying infrared flux increases at MJD 59274 are also presented. Conclusively, cand. I is suggested as a suggestive neutrino emitter.

\section{Discussions and Summary} \label{sec:discu}

It is interesting to put some general theoretical constraints on neutrino production of cand. I based on the multimessenger data. The number of muon (antimuon) neutrinos detected by IceCube during a time interval $ \Delta{T}$ at a declination $ {\delta}$ is given as,

\begin{equation}
     {N_{\nu_{\mu}}}  = \int_{\epsilon_{\nu_{\mu, \text{min}}}}^{\epsilon_{\nu_{\mu, \text{max}}}} {A_{\text{eff}}(\epsilon_{\nu_{\mu}}, \delta) \phi_{\nu_{\mu}} \Delta{T}} \, d\epsilon_{\nu_{\mu}},
\end{equation}

where ${\epsilon_{\nu_{\mu, min}}}$ = 80 TeV and ${\epsilon_{\nu_{\mu, max}}}$ = 8 PeV represent the limits of the energy range that one expects 90$\%$ of neutrinos in the GFU channel \citep{2021JCAP...10..082O}, while ${\phi_{\nu_{\mu}}}$ denotes the differential flux of muon neutrinos. For IC-201221A, the {\tt GFU\_Gold} effective area of IceCube is approximately ${A_{eff} \approx}$ 6 m$^2$ \citep{2023ApJS..269...25A}. Neutrinos are assumed to follow a power-law spectrum ($ {\epsilon}^{-\gamma}$ with $ \gamma$ = 2). Corresponding to the duration of high $\gamma$-ray flux state obtained by the Bayesian Blocks approach, $\Delta{T}$ is set to 2 years. Thus, the integrated muon neutrino energy flux is derived as $\sim$ 1.6 $\times$ 10$ ^{-10}$ erg cm$^{-2}$ s$^{-1}$, corresponding to the detection of one neutrino. Taking into account the high redshift of the blazar, the integrated muon neutrino luminosity is as high as ${\mathcal{L}}_{\nu_\mu}$ $\sim$ 10$^{49}$ erg s$^{-1}$, corresponding to an averaged muon neutrino luminosity ${\epsilon_{\nu_\mu}}{L_{\epsilon_{\nu_\mu}}} =$ ${\mathcal{L}}_{\nu_\mu}$/In(8 PeV/80 TeV) $\sim$ 2 $\times$ 10$^{48}$ erg s$^{-1}$. Alternatively, $\Delta{T}$ is set to 14 years, representative of time length of the entire Fermi-LAT dataset. In this case, the integrated muon neutrino energy flux is reduced to 2.2 $\times$ 10$ ^{-11}$ erg cm$^{-2}$ s$^{-1}$, and so as the averaged muon neutrino luminosity, $\sim$ 3 $\times$ 10$^{47}$ erg s$^{-1}$.

Photopion (${p + \gamma \longrightarrow p + \pi}$) processes are adopted as the responsible neutrino production mechanism, considering the luminous accretion disk emission of the blazar. Energy of the detected neutrino, ${\epsilon_{\nu}}$ $\simeq$ 0.6~PeV in the cosmic rest frame, suggests that energy of the emitting proton is ${\epsilon_p}$ $ {\approx}$ $ {20\epsilon_{\nu}}$ $ {\approx}$ 12 PeV. Quantities with subscripts  `obs' refer to the observers frame, while those with primes denote the frame comoving with the jet, whereas other quantities are measured in the cosmic rest frame , unless specified otherwise. Following estimates in \cite{2018ApJ...865..124M}, ${\epsilon_{\nu}}$ ${\approx}$ 0.6~PeV (0.1 keV/${\epsilon}^{\prime}_t$)($\delta$/10). Considering that for FSRQs external photon fields  are likely responsible to the target photons of photopion processes, the corresponding energy in the cosmic rest frame is inferred, ${\epsilon}_t$ = ${\epsilon}^{\prime}_t/\delta$ $\sim$ 10 eV, where ${\epsilon}^{\prime}_t$ is the energy in the jet comoving frame. The UV emissions could be radiations from BLR. During the interaction, protons lose 3/8 of their energy to neutrinos, resulting in an all-flavor neutrino luminosity given by ${\epsilon_\nu}{L_{\epsilon_{\nu}}} = (3/8){f_{p\pi}}{\epsilon_{p}}{L_{\epsilon_p}}$, where the optical depth to ${p\pi}$ processes (${f_{p\pi}}$) is used to describe the efficiency of neutrino production. The remaining 5/8ths of the lost proton energy contributes to the production of electrons and pionic $\gamma$-rays. Subsequent electromagnetic cascades start and synchrotron cooling is dominated because of Klein–Nishina suppression in inverse Compton processes. The connection between neutrino radiations and the cascade is given as \citep{2018ApJ...865..124M},

\begin{equation}
    {\epsilon_\nu}{L_{\epsilon_{\nu}}} \approx \frac{6\left(1+Y_{IC}\right)}{5}{\epsilon_\gamma}{L_{\epsilon_{\gamma}}}\vert_{\epsilon_{\text{syn}}^{p\pi}} \approx 8 \times 10^{44} \, \text{erg s}^{-1} \left(\frac{{\epsilon_\gamma}{L_{\epsilon_{\gamma}}}\vert_{\epsilon_{\text{syn}}^{p\pi}}}{7 \times 10^{44}}\right).
\end{equation}

Here, $Y_{IC}$ represents the Compton-Y parameter for pairs from the cascades (typically $\leq 1$,  \citealt{2018ApJ...865..124M}). Optimistically, cascade emission is capable of contributing significantly in the $\gamma$-ray domain. Since majority of the observed high-energy $\gamma$-ray photons are around 2~GeV, ${\epsilon_\gamma}{L_{\epsilon_{\gamma}}}$ $\sim$ 5.3 $\times$ 10$^{46}$ erg s$^{-1}$, at $\epsilon_{syn,obs}$ of 2 GeV, is extracted from the observed $\gamma$-ray spectrum in the 2 yr high-flux state. On the other hand, theoretically, $\epsilon_{syn,obs}^{p\pi}$ is set as the same energy. Assuming $\delta$ = 10 and UV emissions as target photons ($\epsilon^{\prime}_t \approx$ 0.1~keV), $\epsilon_{syn,obs}^{p\pi}$ $\approx$ 2 GeV($B^{\prime}/4$ Gauss)(${\epsilon}_{p}$/12 PeV)$^2$(10/$\delta$)$(3.68/(1 + z))$, in which $B^{\prime}$ = 4 Gauss is taken. Note that magnetic field strength of a few Gauss is typical for FSRQs, suggested by SED modeling studies \citep[e.g.,][]{2010MNRAS.402..497G, 2024A&A...681A.119R}. Thus, a muon neutrino luminosity (anti-muon) of 2.1 $\times$ 10$^{46}$ erg s$^{-1}$ is anticipated. Compared to the luminosity of muon neutrinos inferred by the detection of IC-201221A, the Poisson probability of detecting such a neutrino is $\approx$ 0.01. In the case of $\Delta{T}$ = 14 years, a ${\epsilon_\gamma}{L_{\epsilon_{\gamma}}}$, at $\epsilon_{syn,obs}$ of 2 GeV, is $\sim$ 1.4 $\times$ 10$^{46}$ erg s$^{-1}$ \citep{2023arXiv230712546B}, which gives an expected muon (anti-muon) neutrino luminosity of 5.6 $\times$ 10$^{45}$ erg s$^{-1}$. The corresponding neutrino detection probability is suggested as $\approx$ 0.02. An upward fluctuation (${\sim 2}{\sigma}$) is needed if the neutrino is radiated from the blazar, which is consistent with the value found from TXS 0506+056/IC-170922A \citep[e.g.,][]{2019NatAs...3...88G, 2019MNRAS.483L..12C, 2020ApJ...891..115P}.

The proton luminosity of the jet, derived by neutrino luminosity, is compared with Eddington luminosity of the accretion disk,  $1.3 \times 10^{47}$ erg s$^{-1}$ for a black hole mass of $10^{9} M_{\sun}$. In the case that BLR photons (${\epsilon}_t\sim$ 10 eV) are target photons, the optical depth to $p\pi$ processes is suggested as ${f_{p\pi}} \approx \hat{n}_{BLR}{\sigma}_{p\pi}^{eff}r_{BLR} \approx 5.4 \times 10^{-2} f_{COV,-1} L^{1/2}_{AD,46.5}$ \citep{2014PhRvD..90b3007M}. $\hat{n}_{BLR}$ is the number density of BLR photons in the AGN frame, $r_{BLR}$ stands for the BLR radius, $f_{COV} = 0.1$ represents the covering factor \citep{2008MNRAS.387.1669G}, and, ${\sigma}_{p\pi}^{eff}$ $\approx$ $0.7 \times 10^{-28}$ cm$^2$ is the effective cross section \citep{2016PhRvL.116g1101M}. Meanwhile, the accretion-disk luminosity of NVSS J171822+423948 is obtained as $L_{AD}$ = $4.28 \times 10^{46}$ erg s$^{-1}$. Thus, ${f_{p\pi}} \approx 6.3 \times 10^{-2}$ is given. For $\Delta{T}$ = 2 years, ${\epsilon_{p}}{L_{\epsilon_p}}$ = ${\epsilon_{\nu}}{L_{\epsilon_\nu}}$/(3${f_{p\pi}}$/8) $\approx$ $2.7 \times 10^{48}$ erg s$^{-1}$. Assuming a power-law spectrum for protons (${\epsilon_{p}}^{-\gamma}$ with $ \gamma$ = 2, $\epsilon_{p,min} = \delta{m}_p{c}^{2}$ and $\epsilon_{p,max}$ = $10^{18}$ eV, where $m_p{c}^{2}$ is the proton rest energy), as well as $\delta$ =10, ${\mathcal{L}}_{p}$ = ${\epsilon_{p}}{L_{\epsilon_{p}}}$In(10$^{18}$ eV/$\delta{m}_p{c}^{2}$) $\approx$ $5 \times 10^{49}$ erg s$^{-1}$. Therefore, in the AGN frame, the proton luminosity of the jet $\hat{\mathcal{L}}_{p,jet} \approx {\mathcal{L}}_{p} /(4 {\Gamma}^{2}/3)$ $\approx$ $3.8 \times 10^{47}$ erg s$^{-1}$ \citep{2023MNRAS.519.1396S}, with a set of $\Gamma$ = 10. Such a value is approximately 3 times higher than the Eddington luminosity. The required proton power exceeds the Eddington luminosity, which is consistent with the recent studies on the neutrino candidates \citep[e.g.,][]{2019MNRAS.483L..12C, 2019NatAs...3...88G, 2021ApJ...912...54R}. Alternatively, if $\Delta{T}$ = 14 years, the corresponding proton luminosity of jet, $\hat{\mathcal{L}}_{p,jet}$ is $\sim$ 9.9 $\times$ 10$^{46}$ erg s$^{-1}$, which is comparable to the Eddington luminosity.

\begin{figure*}
    \centering
    \includegraphics[scale=0.2]{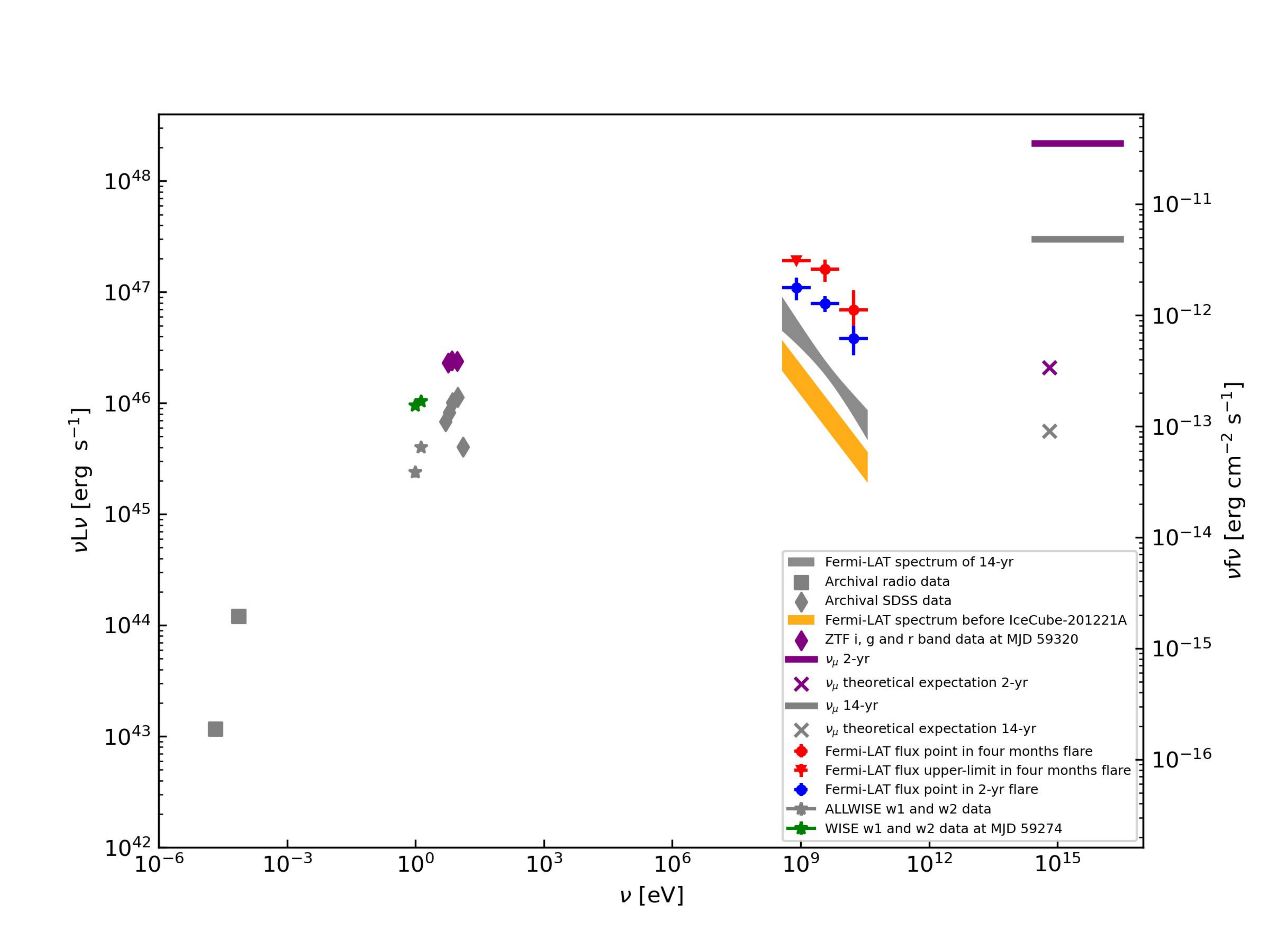}
    \caption {The broadband SED of cand. I. The red, green, and purple points represent data in the four-month flare state, while the blue points correspond to data in the entire two-year flare state. The orange parallelogram represents the Fermi-LAT data before IceCube-201221A, while the grey shadowed butterfly is for the 14-yr Fermi-LAT spectrum. The horizon lines correspond to energy averaged neutrino energy flux for different epochs. Meanwhile, the X-shaped symbols are theoretical anticipations of neutrinos.}
    \label{sed}
\end{figure*}

Besides TXS 0506+056, an intermediate synchrotron peak BL Lac (ISP, \citealt{1995ApJ...444..567P,2010ApJ...716...30A}), several other BL Lacs have been reported as neutrino-emitting blazar candidates, including GB6 J1040+0617 (LSP, \citealt{2019ApJ...880..103G}), MG3 J225517 + 2409, and PKS 0735 + 178 (ISP, \citealt{2020ApJ...893..162F,2023MNRAS.519.1396S}), as well as 3HSP J095507.9+355101 (HSP, \citealt{2020A&A...640L...4G}). Note that an accompanying X-ray flare of 3HSP J095507.9+355101, not in $\gamma$ rays, is detected responsible for the arrival of a neutrino \citep{2020A&A...640L...4G}. FSRQs are also proposed as potential neutrino emitters, such as PKS B1424-418 and GB6 J2113+1121 \citep{2016NatPh..12..807K,2022ApJ...932L..25L}. Interestingly, a minor $\gamma$-ray flare of 1H 0323 + 342, an HSP radio-loud narrow line Seyfert galaxy, is detected temporally coincident with a cospatial neutrino \citep{2020ApJ...893..162F}. Because of the strong optical emission lines, cand. I is classified as an FSRQ. Meanwhile, the shape of its broadband SED suggests that it is an LSP, see Figure \ref{sed}. Accretions of the central SMBHs of FSRQs and RLNLS1s are efficient, leading to intense radiation fields external to the jet. If such photons act as target photons in photo-pion processes, rather than self-radiation from the jet alone, production of neutrinos could be significantly enhanced. On the other hand, dense soft photons also likely play an important role in $\gamma\gamma$ absorption processes, and hence severe absorption creates a soft $\gamma$ -ray spectrum and weakens the observational connection between $\gamma$-ray emissions and neutrinos \citep{2021ApJ...911L..18K}. For LSP/ISP BL Lacs, modest emissions from the accretion system are anticipated, and serious dilution from the jet emission is an obstacle to detect potential broad line emissions \citep[e.g.,][]{2019MNRAS.484L.104P}.

In summary, a thorough investigation of 4FGL J1718.5+4237 has been performed, which is the unique known $\gamma$-ray source within the angular uncertainty region of IC-201221A. First, the low-energy counterpart of the $\gamma$-ray source is determined. Although two radio sources fall into its localization radius, contemporaneous brightening of cand. \uppercase\expandafter{\romannumeral1} (i.e., NVSS J171822+423948) between $\gamma$-ray and optical/infrared emissions are detected. Meanwhile, no optical and infrared sources are found towards the other radio source. Therefore, Cand. \uppercase\expandafter{\romannumeral1} is suggested to be associated with 4FGL J1718.5+4237. Moreover, optical spectroscopic observation suggests that cand. \uppercase\expandafter{\romannumeral1} is a rare high redshift ($z$ = 2.68 $\pm$ 0.01) $\gamma$-ray emitting blazar. The multiwavelength temporal information is also important to establish a potential physical connection between the neutrino and the blazar. It remains in a quiescent state for 12 years undetected by Fermi-LAT, but soon (i.e., a few tens of days) after arrival of the neutrino, a 10-fold $\gamma$-ray flux increase emerges. 115 days later of arrival time of the neutrino, strong optical flares in ZTF $g$, $r$ and $i$ bands are exhibited, meanwhile, infrared fluxes are at high level then. Monte-Carlo simulations, with both spatial and temporal information embraced, give a by-chance probability of $\sim$ 0.03, suggesting that the blazar is a possible neutrino emitter and the first candidate above redshift of 2. Theoretical constraints of neutrino production and comparisons with other candidates have been discussed. Future multiwavelength campaigns, especially X-ray observations in which contributions of cascades of Bethe-Heitler pairs may be non-negligible, together with detailed theoretical SED modeling, would further investigate its jet properties and connection to the neutrino.

\begin{acknowledgments}
We appreciate the instructive suggestions from the anonymous referee that led to a substantial improvement of this work. This research has made use of data obtained from the High Energy Astrophysics Science Archive Research Center (HEASARC), provided by NASA's Goddard Space Flight Center. This research has made use of the NASA/IPAC Infrared Science Archive, which is funded by the NASA and operated by the California Institute of Technology. This study used data based on observations obtained with the Samuel Oschin 48 inch Telescope at the Palomar Observatory as part of the iPTF and ZTF projects. ZTF is supported by the National Science Foundation under grant No. AST-1440341 and a collaboration including Caltech, IPAC, the Weizmann Institute for Science, the Oskar Klein Center at Stockholm University, the University of Maryland, the University of Washington, Deutsches Elektronen-Synchrotron and Humboldt University, Los Alamos National Laboratories, the TANGO Consortium of Taiwan, the University of Wisconsin at Milwaukee, and Lawrence Berkeley National Laboratories. Operations are conducted by COO, IPAC, and UW. This research uses data products from the Wide-field Infrared Survey Explorer, which is a joint project of the University of California, Los Angeles, and the Jet Propulsion Laboratory/California Institute of Technology, funded by the National Aeronautics and Space Administration. This research also makes use of data products from NEOWISE-R, which is a project of the Jet Propulsion Laboratory/California Institute of Technology, funded by the Planetary Science Division of the National Aeronautics and Space Administration.

This work was supported in part by the NSFC under grants 11703093, U2031120, 12192221, 12393814. This work was also supported in part by the Special Natural Science Fund of Guizhou University (grant No. 201911A) and the First-class Physics Promotion Programme (2019) of Guizhou University. 
\end{acknowledgments}

\bibliography{refs}{}
\bibliographystyle{aasjournal}

\end{document}